\documentclass[12pt,a4paper]{article}
\usepackage{amsmath,amssymb}
\usepackage{graphics}
\usepackage{epsfig}
\usepackage{graphicx}
\newcommand{\be}{\begin{equation}}
\newcommand{\ee}{\end{equation}}
\newcommand{\bea}{\begin{eqnarray}}
\newcommand{\eea}{\end{eqnarray}}
\newcommand{\beas}{\begin{eqnarray*}}
\newcommand{\eeas}{\end{eqnarray*}}
\newcommand{\ba}{\begin{array}}
\newcommand{\ea}{\end{array}}
\newcommand{\nn}{\nonumber}

\newcommand{\bt}{\begin{table}}

\newcommand{\al}{\alpha}
\newcommand{\ga}{\gamma}

\newcommand{\de}{\delta}
\newcommand{\De}{\Delta}

\newcommand{\ka}{\kappa}
\newcommand{\la}{\lambda}
\newcommand{\La}{\Lambda}
\newcommand{\na}{\nabla}

\newcommand{\vphi}{\varphi}
\newcommand{\si}{\sigma}

\begin{document}
\title{
{\bf 
Quartet-metric gravity,  scalar-graviton dark substance
and vacuum energy screening: 
extending GR vs.\ its  WTDiff alternative 
}}
\author{Yury~F.~Pirogov
\\
{\small
Theory Division, Institute for High Energy Physics of   NRC  Kurchatov Institute, }\\
{\small
Protvino, 142281 Moscow Region, Russia
}
}
\date{}
\maketitle
\begin{abstract}
\noindent
In the frameworks of the  effective field theory 
of  metric supplemented by  some distinct dynamical coordinates  
parametrized, in turn,  by  a scalar quartet
-- the so-called quartet-metric gravity -- 
the extension of  tensor gravity through a massive scalar graviton 
in addition to the  massless tensor one is  consistently exposed. 
The field equations for the  two  realizations of such 
an extension originating from the classically equivalent prototype theories --
General Relativity  (GR) and  its Weyl transverse (WTDiff) alternative -- 
are derived and argued to be, generally, non-equivalent,
with the pure-gravity case manifesting this explicitly in detail.
A splitting  of the cosmological constant onto the gravitating 
and non-gravitating parts, with a  partial screening of the  vacuum energy 
through  an emergent scalar-graviton dark  substance, is considered.
A prior importance  of treating  the WTDiff gravity as a prototype one on par with GR, 
when looking for  a putative next-to-GR extended theory of gravity with a  scalar-graviton 
dark substance,  is stressed.\\



\end{abstract}

\section{Introduction: beyond GR through quartet modification}
\label{intro}

The contemporary   Cosmological Standard Model, or, otherwise, the $\La$CDM model  
accumulates the state-of-the-art for the present-day description of the evolution of the 
Universe.\footnote{For a concise exposition  of all the relevant topics,  see~\cite{PDG}.}
According to its very  name, the model incorporates such  new ingredients  of 
cosmology as the cosmological constant (CC)
$\La$ comprising at the present epoch about 70$\%$ of the  partial energy density  
of the Universe  
and a  (cold) dark matter (DM) comprising about 25$\%$ of such an ``energy budget''.
At that, DM serves as  a corner-stone for building the dark halos 
of the galaxies and  cluster of galaxies.
Being extremely economic
in  its basic concepts,  such a model  nevertheless shows an impressive
success in describing the wide variety of the observational data.
Still, some arguments (predominantly of the  theoretical origin) 
may imply a necessity of  going eventually 
beyond $\La$CDM.\footnote{For $\La$CDM 
and beyond, see, e.g.,~\cite{Joyce, Bull}.}  

One of such an arguments is provided by the so-called 
vacuum energy/CC problem which may be   
at least threefold.\footnote{For the vacuum energy/CC
(or, more generally,  dark energy (DE))  problem, see, e.g,~\cite{Wein}--\cite{Burg}.}   
First of all, why the CC $\La$,  though being rather large on the cosmological scale,   
is  nevertheless  unnaturally small compared to what might be expected in 
General Relativity (GR) as the effective field theory (EFT)?
Secondly, why the classical $\La$, being ones put relatively small, still  remains stable 
with respect to  the  quantum corrections? And at last, 
why $\La$ starts to manifest  itself only at the rather late cosmological epoch 
(or,  as it  is sometimes stated,  ``why now'')? 
Though causing no principle difficulties phenomenologically, 
the CC problem  may present  theoretically the greatest challenge 
to the fundamental physics.
Another  crucial problem  for  the gravity 
and cosmology is the DM one:\footnote{For a cumulative view on DM, cf, e.g.,~\cite{PDG}.}  
what is the real nature of DM, especially 
in relation with  the Particle Standard Model?
Being not as  principle as CC for the theoretical consistency of GR, 
the (cold) DM  causes still  definite  tension within the GR approach.
Thus, though GR and the built on it $\La$CDM are up to now  in a rather solid shape, 
nevertheless  some their modifications/extensions may 
be in order.\footnote{For the modified and extended theories  of gravity
beyond GR, cf., e.g.,~\cite{Capo, Clift}.}
Moreover, the DE and DM problems  may 
even be the heralds  of the future crucial changes in the  present-day 
paradigms for   gravity and cosmology.\footnote{For importance of 
the  new paradigms when accumulating the problems 
within the old ones, cf., e.g., ~\cite{Kuhn}.}

In this vein, in refs.~\cite{Pir1}--\cite{Pir3} there was proposed  
EFT of the so-called {\em quartet-modified/quartet-metric} gravity,
the latter  being based on the three following  
physical concepts.\footnote{Such a gravity modification  
is taken here ad hoc as EFT on its own. Nevertheless, it admits a  justification 
and partial refinement in 
the frameworks of a more fundamental  {\em affine-Goldstone} nonlinear model 
defined at and proliferated below the Planck scale~\cite{Pir4}.}
First, in addition to a dynamical tensor field/``bare'' metric  
there exist in spacetime  some distinct dynamical coordinates 
(to be associated ultimately with  the vacuum)
defined by a scalar quartet.\footnote{This concerns  the four spacetime dimensions, 
with the proliferation to an arbitrary case being straightforward.}  
Second, such a scalar quartet plays  for gravity the role of the Higgs-like field, 
through  absorbing  (a part of) the components of which 
a part of the (formerly  gauge) components  in  metric gets physical. 
Third,  the additional physical gravity components serve as an emergent dark substance 
(DM, DE, etc.) of the Universe. 
The respective EFT of the quartet-metric gravity  is thus basically defined 
on the extended set of the fourteen fields and 
is  invariant under the four-parameter general diffeomorphism (GDiff) symmetry
leaving  at most the ten independent gravity components. 
A priori, the theory 
describes in a  completely dynamical, 
GDiff invariant and  generally-covariant (GC)  fashion 
the (in general, massive)  tensor, scalar and  vector  ``gravitons`` 
(a part of which  may, in fact, be unphysical).
By this token, the mere admixture to metric of the scalar quartet may 
result  in  an extremely  rich spectrum of the emergent   physical phenomena beyond GR,  
described by  a wide  variety of the  particular realizations 
of  the generic quartet-metric gravity.  

To tame the ensuing ambiguities, one may  adhere to  
the scalar-reduced quartet-metric gravity,   with only a massive scalar graviton  
in addition to  the conventional  (massless) tensor one of~GR,  
as the most simple and natural version of the extended 
next-to-GR theory of gravity within the quartet-modified frameworks.\footnote{The earlier   
extension to GR~\cite{Pir5, Pir6}, with  the scalar graviton 
containing a non-dynamical scalar density to  reconstruct GC, 
may naturally be embodied  in the completely dynamical frameworks of the present paper.}
But  even if one adopts  the concept of the scalar graviton, there still remains 
the wide residual freedom  in choosing a preferred mode of its particular realization.
Besides the evident ambiguity in choosing a particular 
Lagrangian for the scalar-graviton extension,
there is also left an ambiguity  in choosing 
the prototype/``graft''  theory of gravity undergoing such a   scalar-graviton extension. 
Namely, one may choose either 
(i) GR  with the GDiff invariance, or
(ii) the Weyl transverse/WTDiff gravity  which,   
though  restricted by the transverse Diff's (TDiff's),  still allows one   
more gauge transformation -- the local scale  one/Weyl 
rescaling.\footnote{For the WTDiff gravity as a viable  alternative to GR,  
see, e.g.,~\cite{Iza}--\cite{Oda}.}  
The two prototype theories -- GR and WTDiff gravity -- 
prove to be  classically equivalent  
(under the covariant conservation 
of the energy-momentum tensor), both  
containing  the ten  dynamical metric components undergoing 
the four-parameter gauge transformations GDiff or WTDiff, respectively. 
The latter ones leave   
the six physical metric components off-mass-shell describing  the two-component 
massless tensor graviton on-mass-shell.\footnote{For comparison,
the unimodular gravity, with  only nine  dynamical metric components   
(the determinant of the metric  being a priori fixed, say,  to unity) undergoing   
the three-parameter gauge symmetry TDiff,
results in  the same count for the massless tensor graviton getting classically 
equivalent to GR and, thus, to the WTDiff gravity (modulo a cosmological constant).} 
What concerns CC, in GR the quantum CC problem is inborn 
as a manifestation of  the longitudinal gravity component off-mass-shell.
On the contrary, in the WTDiff gravity  the  Lagrangian CC  
gets, by the very construction, 
irrelevant, being   substituted by an  integration constant. The latter 
proves to be stable due to the Weyl invariance 
against the quantum corrections~\cite{Iza, Carb3, Alv4}. 
Nevertheless,  though  equivalent classically as the prototype  theories, GR and the 
WTDiff gravity may  result in the  
non-equivalent extended  theories  already on the classical (moreover, on  quantum) level. 
With  this  in mind, we consider  in the present paper  the two aforementioned  alternatives  
a priori on par to future choosing  the most relevant one  
in the context of the emergent scalar-graviton 
dark substance and  the CC/vacuum energy problem.

In Section~2, the basics  of the quartet-metric gravity as  EFT is recapitulated, 
with a generic splitting of CC onto the two parts -- gravitating and non-gravitating -- 
emphasized.
In Section~3, the consistent scalar-graviton reduction of the  generic quartet-metric 
gravity  is presented starting from the first principles of the latter.
A more restrictive model  interpolating between GR and WTDiff gravity 
as the classically equivalent prototype theories, 
which  result in the two non-equivalent extended theories,
is  presented. The gravitational gauge symmetry differing these marginal cases is studied 
for  both the dynamical and non-dynamical scalar densities 
entering the  (composite) scalar-graviton field.
In Section~4, such a  scalar-graviton reduction 
is studied more particularly for GR as a prototype theory, 
while  in Section~5 the same is done for its WTDiff alternative, with the non-equivalence 
of the two extensions being shown explicitly  in the pure-gravity/matterless  case.
In Summary, the two alternatives are compared in respect to  
the  emergent scalar-graviton dark substance and the  
(partial) screening of the  proper vacuum energy. 
The conceivable advantages of the WTDiff extension are advocated, and 
a prior  necessity to account for both alternatives, 
when going beyond GR  to  ultimately choosing from them the most 
relevant (if any) as the prototype one   for the next-to-GR  
scalar-graviton extended theory  of gravity,  is emphasized.

\section{Quartet-modified/quartet-metric  gravity: generalities}
\label{sec:2}

The EFT  of the quartet-metric gravity~\cite{Pir1}--\cite{Pir3} 
is generically defined  by  a  GC scalar action functional
\be\label{S}
S[g_{\mu\nu},  {{Q}}^a]  = \int {\cal L}_G(g_{\mu\nu},   {{Q}}^a) \, d^4 x ,
\ee
with  a  Lagrangian scalar density ${\cal  L}_G$ 
dependent on the two basic fields as the functions 
of the arbitrary observer's/kinematic coordinates 
$x^\mu$ ($\mu=0,\dots,3$): a symmetric  tensor  $g_{\mu\nu}(x)$  
and  a quartet of the  scalar fields  
${{Q}}^{a}(x)$, $a=0,\dots,3$.
More particularly, let    $a,b,\dots$ be the  indices of the global   Lorentz group  of 
the reparametrizations  ${{Q}}^a\to \La^a{}_b {{Q}}^b$, $\La\in SO(1,3)$,  
possessing the  invariant  Minkowski symbol 
$\eta_{ab}$.  By default, the
signatures of $g_{\mu\nu}$ and $\eta_{ab}$ are  chosen  to  coincide.  
Assume moreover that  the
scalar fields ${{Q}}^{a}$ admit the 
global (not related to  the spacetime) Poincar\'e reparametrizations 
composed of the Lorentz ones  and the shifts ${{{Q}}}^{a}\to {{{Q}}}^{a}+C^{a}$, 
with the  arbitrary
constant parameters~$C^{a}$.
Due to the global Poincar\'e invariance, 
${{Q}}^a$ should, in fact, enter the action
through an auxiliary  {\em quasi-affine} metric  
\be
{Q}_{\mu\nu}\equiv\partial_\mu  {{{Q}}}^a  \partial_\nu {{{Q}}}^b \eta_{ab}
\ee
and
\be\label{key}
{{\cal Q}}\equiv \det({{Q}}_{\mu\nu})  = \det (\partial_\mu 
{{{Q}}}^a)^2 \det(\eta_{ab})<0.
\ee
For  the  non-degeneracy of the quasi-affine 
metric ${{Q}}_{\mu\nu}$, possessing thus  an inverse ${{Q}}^{-1\mu\nu}$,
there should fulfill ${\cal Q}\neq 0$ or $\infty$.
Consider now a maximal connected spacetime region (an ``affine patch'') 
where the Jacobian 
$J=\det(\partial {{Q}}^a/\partial x^\mu)\neq 0$ or $\infty$
allowing thus to invert 
the dependence ${{Q}}^a={{Q}}^a(x)$   to $x^\mu= x^\mu({{Q}})$. 
By this token, we can  cover the spacetime manifold $M_4$ in a patch-wise fashion
by  some  distinct dynamical coordinates -- the {\em quasi-affine} ones -- 
${\mathring x}^{\al}=\de^\al_a {{Q}}^{a}(x)$, $\al=0,\dots, 3$,  
with  an inverse $x^\mu=x^\mu({\mathring x})$.\footnote{The edges
of the  affine patches and the  singular  points (if any),
where the invertibility of ${\mathring x}^\al= {\mathring x}^\al(x)$ 
breaks down, are to be treated separately.} 
Operationally, the  quasi-affine coordinates ${\mathring x}^\al$   
are distinct by the fact that under  using them 
the quasi-affine metric gets  Minkowskian form,  
${{Q}}_{\al\beta}({\mathring x})\equiv \eta_{\al\beta}$ 
(respectively, ${{Q}}^{-1 \al\beta}({\mathring x})\equiv \eta^{\al\beta}$).\footnote{At that, 
generally, $g_{\al\beta}({\mathring x}) \neq \eta_{\al\beta}$. And v.v.,
in the locally inertial coordinates in a vicinity of a point, where $g_{\mu\nu}$ 
is approximately diagonal, the quasi-affine metric is not bound to be such.}
Physically,  the coordinates ${\mathring x}^\al$  may 
be postulated   as those comoving with the  vacuum, 
the latter  treated ultimately as a dynamical  system on par 
with all the dynamical fields including the bare metric 
$g_{\mu\nu}$.\footnote{At the quantum 
level, similarly to  the decomposition of the metric 
$g_{\mu\nu}=\bar g_{\mu\nu}+h_{\mu\nu}$,  we should put 
$ {{Q}}^a= \bar {{Q}}^a+q^a$ and ${\mathring x}^\al=\bar  {\mathring x} ^\al +\chi^\al$
defining the background coordinates $\bar {\mathring x}^\al \equiv \de_a^\al  {\bar {Q}}^a $ 
and  some small quantum fluctuations $h_{\mu\nu}$,   $q^a$ and $\chi^\al=\de^\al_a q^a$ 
relative to  the backgrounds. In essence, 
this may be considered as a definition of a kind of the quantum spacetime.}

Altogether,  the action of  the quartet-metric gravity
may most generally be rewritten  in an equivalent entirely  spacetime GC  form as
\be
S=\int {\cal L}_{G}(g_{\mu\nu}, {{Q}}_{\mu\nu}, g, {{\cal Q}}) d^4 x ,
\ee
where $g\equiv \det (g_{\mu\nu})<0$.
At that, as the basic dynamical variable there still serves  the  scalar quartet~${{Q}}^a$
(in the line with  the bare metric $g_{\mu\nu}$) in terms of which 
the consideration ultimately proceeds. 
The Lagrangian density ${\cal L}_G$ may further be decomposed as 
\be\label{calL}
{\cal L}_G= L_G(g_{\mu\nu},  {{Q}}_{\mu\nu},g/{\cal Q} ) \, {\cal M}(g, {{{\cal Q}}}),
\ee
with a GC scalar  Lagrangian $L_G $  
supplemented by a spacetime measure ${\cal M}$.  The latter is  a GC scalar density 
of the proper weight entering the  spacetime volume element 
$d V= {\cal M}d^4 x$ to make the latter a true GC scalar.
In view of ${{\cal Q}}\neq 0$, the sign of $\sqrt{-{{\cal Q}}}$ 
is (patch-wise) preserved and we can put 
$\sqrt{-{{\cal Q}}}>0$. 
A priori,  the measure is defined up to 
a scalar function  $ \varphi_{{\cal M}}(g/{{\cal Q}}) $,
which may be attributed, if desired, to $L_G$.
Thus, with the proper  redefinition of $L_G$, the measure may   
equivalently  be chosen either as $\sqrt {-g}$ or $\sqrt {-{{\cal Q}}}$, 
depending on the context.
Altogether, prior to fixing the Lagrangian we can without loss of generality put 
\be
S= \int { L}_G(g_{\mu\nu},  {{Q}}_{\mu\nu},{{\cal Q}}/g ) \sqrt{-g}\, d^4 x 
\ee
(up to $\sqrt{-g}\leftrightarrow \sqrt{-{\cal Q}}$).
In the frameworks of the quartet-metric gravity it is always  possible to include
in  ${\cal L}_{G}$ also  the field-independent pure-measure contribution 
\be\label{DeLLa}
\De{\cal L}_\La= - \ka_g^2(\La_g\sqrt{-g}  +\La_{Q} \sqrt{-{{\cal Q}}}),
\ee
with $\ka_g$   the truncated Planck mass, containing instead of a  single conventional CC $\La$
the two, generally, independent  CC's: 
a ``gravitating''/Riemannian $\La_g$ 
and a ``non-gravitating''/quasi-affine $\La_{Q}$
of the dimension mass squared.
Such a  conceivable CC splitting to account for the vacuum energy 
is a generic  trait of  the quartet-metric gravity compared to GR and its direct 
siblings.\footnote{For a non-gravitating vacuum energy/CC 
associated with a non-geometrical measure  constructed 
from a scalar quartet, see, e.g.,~\cite{Guendel1, Guendel2}.} 

Generally, the  Lagrangian $ L_G$  
describes the multi-component gravity 
mediated  by  the  (massive)  scalar,  
tensor  and vector  gravitons  (a part of which being, conceivably,  unphysical) contained 
in the metric field. At that, the quartet  ${{Q}}^{a}$ serves ultimately as a
gravity counterpart  of the Higgs field which 
provides the four additional independent components.\footnote{More precisely, 
instead of ${Q}_{\mu\nu}$ 
we could equivalently use a Higgs-like field~\cite{Pir4} $H^\mu_\nu=
g^{\mu\la}{Q}_{\la\nu}=g^{\mu\la}\partial_\la Q^a\partial_\nu Q^b\eta_{ab}$ 
(respectively, $H^{-1}{}_\mu^\nu=g_{\mu\la}{Q}^{-1\la\nu}$), 
so that $\det (H^\mu_\nu)={\cal Q}/g$,
with $H^\mu_\nu$  defining 
the derivativeless/potential part of $L_G$ as $V_G(H)
=\sum_n c_n\mbox{\rm tr} (H^n)$, with the arbitrary  coefficients $c_n$ 
at  the  (including negative) degrees $n$ of~$H$. 
Equivalently, as a counterpart of $H^\mu_\nu$ 
there may serve $M^{ab}\equiv g^{\mu\nu}\partial_\mu Q^a\partial_\nu Q^b$.}  
The Lagrangian   $ L_G$ quadratic in the first
derivatives  of metric is constructed  in~\cite{Pir1}.
A more  general  quartet-metric Lagrangian  is discussed in~\cite{Pir2, Pir3}.
Imposing  on the  parameters of $L_{G}$ the 
``natural'' (in a technical sense) 
restrictions we can exclude 
the vector graviton, as the most ``suspicious'', leaving   
in addition to the massless tensor
graviton only the  massive scalar one to be treated in what follows.

\section{Scalar-graviton reduction}   

\subsection{Generic model}  

The quartet-metric gravity  significantly simplifies  (remaining still very
rich of a new content) under the scalar-graviton reduction given by  the 
Lagrangian  which depends on the quasi-affine metric ${{Q}}_{\mu\nu}$
exclusively  through its determinant~${{\cal Q}}$:  
\be
S=\int {\cal L}_{gs}(g_{\mu\nu},g, {{\cal Q}}) d^4 x
\equiv  \int  {\cal L}_{gs}(g_{\mu\nu},{{\cal Q}}/g, g)d^4 x. 
\ee
With  the ratio ${{\cal Q}}/g$ being equivalently substituted by  
\be\label{si}
\si\equiv \ln \sqrt{-g}/\sqrt{-{{\cal Q}}}, 
\ee 
we  can always put, say, 
\be
{\cal L}_{gs}(g_{\mu\nu},g, {{\cal Q}}) = 
 L_{gs}(g_{\mu\nu},\si)\sqrt{-g}.
\ee
Due to ${\cal Q} $  having the same weight  as 
$g$ under the general coordinate  transformations,  
$\si$ is a true GC scalar with the normalization $\si|_{g={{\cal Q}}} =0$. 
Stress that $\si$, to be called the {\em scalar 
graviton},  has a  composite nature, ultimately distinguishing such a scalar field
from the elementary one.
Eqs.~(\ref{si}) and   (\ref{key}), with
$g_{\mu\nu}$ and ${{Q}}^{a}$ as the  independent 
field variables,  are the  key ingredients of the completely 
dynamical  theory of the scalar graviton within the quartet-modified frameworks. 

Constructing such a theory was undertaken in~\cite{Pir3} 
based, in particular, on a general  (matterless)  scalar-tensor
theory (in the four spacetime dimensions)  for 
the (bare) metric/tensor field $g_{\mu\nu}$ supplemented by 
a conventional scalar field~\cite{Berg, Horn}. An important distinction 
for the (composite) scalar graviton stems, though,   
from the   constraint~(\ref{si}).  
One more modification of the theory is still in order. Namely, 
let us introduce the conformally rescaled  metric
\be\label{tilde}
\tilde  g_{\mu\nu}\equiv  \tilde \varphi_g  (\si) g_{\mu\nu},
\ee
with $(-\det (\tilde g_{\mu\nu}))^{1/2}  \equiv\sqrt{-\tilde g}=\tilde\varphi_g^2 \sqrt{-g}$,
and $\tilde  g^{-1\mu\nu}=\tilde\varphi_g^{-1}  (\si) g^{\mu\nu}$ 
being an inverse of $\tilde  g_{\mu\nu}$.
By this token, the most general Lagrangian density 
for the (matterless)  scalar-tensor gravity in the  quartet-metric frameworks 
may be presented equivalently as 
\be\label{Horn'}
 \tilde {\cal L}_{gs} = 
 \tilde L_{gs}(\tilde g_{\mu\nu},\si) \sqrt{-\tilde g} 
\ee
to be generally understood in what follows.\footnote{For the consistent 
geometrical interpretation of the theory,  the spacetime 
tensor indices are assumed to be 
manipulated   by  the effective metric $\tilde  g_{\mu\nu}$  (or $\tilde  g^{-1\mu\nu}$),
but not by the bare one  $g_{\mu\nu}$ 
(or $g^{-1\mu\nu}\equiv g^{\mu\nu}$).}$^,$\footnote{Allowing for
the dependence on the whole ${{Q}}_{\mu\nu}$, 
the  effective metric in the quartet-metric frameworks 
could be taken even more generally as  the {\em ``disformal''} one 
$\tilde  g_{\mu\nu}\equiv  \tilde \varphi_g  (\si) g_{\mu\nu}
+ \tilde \varphi_{Q}  (\si){{Q}}_{\mu\nu}$, with 
the two scalar functions $\tilde\varphi_g  (\si)$ and $\tilde\varphi_{Q}  (\si)$ 
bound to assure the non-degeneracy of $\tilde  g_{\mu\nu}$.} 
Finally, in the presence of matter the  respective Lagrangian density  looks like 
\be\label{Lgsm}
\tilde {\cal L}_{gsm}=\tilde L_{gsm}(\tilde g_{\mu\nu}, \si,\phi_I)\sqrt{-\tilde g},
\ee
where $\phi_I$ is a generic matter field. 
At that, the effective metric  $\tilde g_{\mu\nu}$
remains a priori unspecified  due to the residual conformal redefinitions. 
To abandon such an  ambiguity, there is considered in what follows even more  restrictive 
but still rather general class of  the scalar-tensor theories embodying, 
hopefully, a looked-for next-to-GR one containing the scalar graviton.

\subsection{Factorization  model}  

So, let us further postulate the  Lagrangian  sufficiently  
generally in the partially factorized form 
corresponding to  the effective pure-tensor gravity plus the rest:
\be
\tilde L_{gsm}
= \tilde L_g(\partial_\ka  \tilde g_{\mu\nu}, \tilde g_{\mu\nu})+
\tilde L_{sm}(\partial_\ka\si,\partial_\ka\phi_I, \tilde g_{\mu\nu},\si,\phi_I)
\ee
and subsequently fixing the form of  $\tilde L_g$.
Minimally, we can  put 
\be
 \tilde L_g= -\frac{1}{2}\ka_g^2 R(\tilde g_{\mu\nu}), 
\ee
with $\ka_g=1/\sqrt{8\pi G_N}$ being the truncated  Planck scale, 
$\tilde R \equiv  R(\tilde g_{\mu\nu})   = \tilde g^{-1\ka\la} R_{\ka\la}(\tilde g_{\mu\nu})   
$ the Ricci scalar  and   $R_{\ka\la}(\tilde g_{\mu\nu})$ the   Ricci tensor. 
After  choosing  the pure-tensor gravity Lagrangian $\tilde L_g$ the only 
freedom remains  in  
the scalar-matter Lagrangian $\tilde L_{sm}$, being a priori an arbitrary 
function of its arguments.

More particularly,  picking-up the dependence on the derivatives of the fields, 
we could further assume the  factorization:
\be
\tilde L_{sm}
= \tilde L_s(  \partial_\ka \si,\tilde g_{\mu\nu}, \si,\phi_I)+
\tilde L_m(\partial_\ka \phi_I, \tilde g_{\mu\nu},\phi_I,\si)-V_{sm}(\si,\phi_I),
\ee
with $V_{sm}(\si,\phi_I)$ being the most general GC scalar potential.
In the second-derivative approximation, 
the scalar-graviton Lagrangian  may be taken in  the 
most general quadratic form as 
\be\label{L_s}
\tilde L_{s}=\frac{1}{2}\ka_s^2 \tilde\varphi_s(\si,\phi_I)
\tilde g^{-1\mu\nu}\partial_\mu\si \partial_\nu\si ,
\ee
where $\ka_s\ll \ka_g$ is a scalar-graviton scale. 
If the kinetic  profile function of the scalar graviton $ \tilde\varphi_s$ is independent of $\phi_I$, 
eq.~(\ref{L_s})  implies   a redefinition of the scalar-graviton field through
\be
\si \to  \tilde\si=\int^\si   \tilde\varphi^{1/2}_s (\si') d \si'.
\ee
The kinetic matter Lagrangian $\tilde L_m$ remains still an arbitrary function of its arguments.
The potential $V_{sm}$ 
describes generically 
the masses of the fields and  their interactions  
(incorporating, possibly, the spontaneous symmetry breaking).

\subsection{GR-to-WTDiff interpolation model}

\subsubsection{WGDiff gauge symmetry}

To grasp the essence of the scalar-graviton reduction consider 
even more restrictive but still sufficiently  general  
model given by the effective metric  corresponding to 
$\tilde \varphi_g(\si)=e^{-\ga\si/2}$, which   depends on an arbitrary constant parameter~$\ga$,
so that 
\be
\tilde g_{\mu\nu} \equiv e^{-\ga\si/2} g_{\mu\nu}
= ({{\cal Q}}/g)^{\ga/4} g_{\mu\nu},
\ee
with an inverse $\tilde g^{-1\mu\nu} =e^{\ga\si/2} g^{\mu\nu}
=(g/{{\cal Q}})^{\ga/4} g^{\mu\nu} $ and $\tilde g =  g^{(1-\ga)} {{\cal Q}}^{\ga}$.
At $\ga\neq 0$,  the effective metric $\tilde g_{\mu\nu}$ superficially contains 
the exhaustive number, eleven, of the  independent variables. 
A trait of GR is its gauge symmetry -- GDiff -- on which GR 
(and its direct siblings) are, in essence, grounded.
So, a  concise way of treating a modified gravity beyond GR 
is to  consider the modification  of the respective  gauge symmetry 
compared to the conventional GDiff taken as the reference one.
This could, generally, allow to 
reduce the  number  of the independent variables  
in the effective metric $\tilde g_{\mu\nu}$ to the conventional six. To this end,  
let us first introduce for the 
bare metric $g_{\mu\nu}$ 
the  five-parameter combined gauge symmetry 
consisting of the  GDiff  transformations -- the Lie derivatives --  
defined  by  a   vector field $\xi^\la(x)$
supplemented by the Weyl rescaling transformations defined  by a scalar field $\zeta(x)$.
In these terms, an infinitesimal gauge transformation
$D$  -- a combined Lie derivative -- is acting on the bare metric  
$g_{\mu\nu}$ as follows:
\bea
{D}  g_{\mu\nu}&=& g_{\mu\la}\partial_\nu  \xi^\la 
+ g_{\nu\la} \partial_\mu \xi^\la+  
\xi^\la \partial_\la g_{\mu\nu}+ \zeta   g_{\mu\nu} \nn\\
&=& g_{\mu\la}\na_\nu \xi^\la+ g_{\nu\la}\na_\mu \xi^\la +\zeta   g_{\mu\nu}, \nn\\
{ D} \sqrt{-g}/\sqrt{-g}&=& 1/2\,  g^{\mu\nu} { D} g_{\mu\nu} =
\partial_\la (\sqrt{-g}\xi^\la)/\sqrt{-g}  +2\zeta  = \na_\la\xi^\la +2\zeta  ,
\eea
with $\na_\la$ being a covariant derivative with respect to  $g_{\mu\nu}$. 
Call the respective  symmetry the {\em WGDiff} one.
The transformation of $\cal Q$ with respect to WGDiff depends on whether $\cal Q$   
is dynamical or not.

\subsubsection{Non-dynamical scalar density}

To better elucidate the role of $\cal Q$ as  dynamical, let first  it be a priori   
given  non-dynamical/``abso\-lute''  scalar density, still 
appropriate  for dealing with  the WTDiff gravity and the scalar graviton.
To effectively ``freeze'' $\cal Q$ consider the restricted infinitesimal 
gauge transformation $\mathring {D}$ 
putting  by default  $  \mathring {D}{Q^a}\equiv 0$ with $\mathring {D}{\cal Q}=0$, but 
leaving still $\mathring{D}  g_{\mu\nu}= {D}  g_{\mu\nu}$ as indicated above, so that 
\bea\label{Dsi}
\mathring{ D} \si&=& \mathring{ D} \sqrt{-g}/\sqrt{-g}= 1/2\,  g^{\mu\nu} { D} g_{\mu\nu}
=  \partial_\la (\sqrt{-g}\xi^\la)/\sqrt{-g}  +2\zeta  \nn\\ 
&=& 
\xi^\la\partial_\la\si+   \partial_\la (\sqrt{-{\cal Q}}\xi^\la)/\sqrt{-{\cal Q}}  +2\zeta,
\eea
with the ensuing
\be\label{ga2}
\mathring { D} \tilde  g_{\mu\nu}
= \tilde g_{\mu\la} \partial_\nu\xi^\la  
+ \tilde g_{\nu\la} \partial_\mu\xi^\la 
+\xi^\la\partial_\la \tilde g_{\mu\nu} -\ga/2\, 
\partial_\la (\sqrt{-{\cal Q}}\xi^\la)/\sqrt{-{\cal Q}}\, \tilde  g_{\mu\nu}
+(1-\ga)\zeta \tilde  g_{\mu\nu} .
\ee
It follows  hereof that the invariance of  
the Lagrangian $\tilde L_g\sim R(\tilde  g_{\mu\nu}) $  at an arbitrary $\ga$ 
forbids a gauge transformation given by a
$\ga$-dependent combination of the  longitudinal (putting for definiteness ${\cal Q}=-1$)
Diff  and the Weyl rescaling.
In particular, at $\ga=0$
this implies  the contraction to $\zeta\equiv 0$ under  an arbitrary $\xi^\la$,
what in turn implies the residual GDiff. On the other hand,  at $\ga=1$  
there follows the contraction to the transversal $\xi^\la$ satisfying to 
$\partial_\la \xi^\la\equiv 0$ under an arbitrary $\zeta$, 
implying the residual WTDiff.
At the intermediate $0<\ga<1$, there takes place   a
four-parameter gauge symmetry  in-between GDiff and WTDiff.\footnote{At a fixed 
non-dynamical ${\cal Q}$ (say, ${\cal Q}=-1$) the 
two marginal cases of the interpolating model for  $\ga=0$ and $\ga=1$  
prove  to be  the only ones 
describing the two-component massless tensor graviton 
in terms of the  ten-component metric field, 
with the difference between the cases  corresponding to   
the required four-parameter gauge symmetry: GDiff vs.\ WTDiff~\cite{Alv1}.} 
At that, it follows from (\ref{Dsi}) that the requirement for  $\si$    
to transform as  a  scalar contracts  the gauge symmetry  in an extended  Lagrangian   
to  $\partial_\la\xi^\la=\zeta\equiv 0$ 
implying the residual gauge symmetry to be TDiff in any case,
whether GDiff ($\ga=0$) or WTDiff ($\ga=1$). The  TDiff symmetry proves to be  
precisely what signifies at a non-dynamical $\cal Q$  the appearance of a scalar graviton 
in excess of the massless tensor one.\footnote{In fact, 
the terms with the derivatives of $\si$ remain still invariant 
under the  global shift symmetry $\si\to \si+C$. 
This symmetry may be imposed to suppress the derivativeless dependence on  
$\si$ on its own.}$^,$\footnote{
Under a non-dynamical~${\cal Q}$, to the previous list 
of the gauge symmetries there may be  added 
the three-parameter TDiff for the nine independent metric variables, 
with $g/{\cal Q}\equiv 1$ implying  an elementary scalar field as a
counterpart of the scalar graviton to serve  as a dark substance. 
This is in contrast to 
the dynamical~$\cal Q$, 
with the scalar graviton being, in fact,  composed of the two  fields,  $\si=\si(g/{\cal Q})$, 
what determines many peculiarities of the latter.}

\subsubsection{Dynamical scalar density}

For a dynamical $\cal Q$, proliferate  $D$ further on ${Q}_{\mu\nu}$ 
as a conventional Lie derivative\footnote{The  
local gauge transformation  
$D{{Q}}^a\sim \zeta{{Q}}^a$ would violate the assumed global shift symmetry for ${{Q}}^a$,
and so is not admitted.}
\bea\label{D}
{ D} {{Q}}^a&=&\xi^\la \partial_\la {{Q}}^a\equiv \xi^\la  {{Q}}^a_\la,\nn\\
{ D} {{Q}}^a_\mu =\partial_\mu( {D} {{Q}}^a)
&=&{{Q}}^a_\la\partial_\mu \xi^\la+ \xi^\la \partial_\la {{Q}}^a_\mu,\nn\\
{ D}  {Q}_{\mu\nu}&=&  \eta_{ab}({{Q}}^a_\mu { D}  {{Q}}^b_\nu
+{{Q}}^{a}_\nu { D} {{Q}}^b_\mu) \nn\\   
&=&  {Q}_{\mu\la}\partial_\nu  \xi^\la 
+ {Q}_{\nu\la} \partial_\mu \xi^\la+ 
\xi^\la \partial_\la  {Q}_{\mu\nu},
\eea
with the  use being  made of 
$ \partial_\mu {{Q}}^a_\la =\partial_\mu\partial_\la {{Q}}^a =\partial_\la {{Q}}^a_\mu$, 
so that 
\bea\label{calD}
{ D} \sqrt{-{\cal Q}}/\sqrt{-{\cal Q}}=1/2\, {Q}^{-1\mu\nu} { D} {Q}_{\mu\nu }  
&=&  \partial_\la (\sqrt{-{\cal Q}}\xi^\la)/\sqrt{-{\cal Q}}=  
\xi^\la  \partial_\la \ln \sqrt{-{\cal Q}} + \partial_\la\xi^\la. 
\eea
This results in the infinitesimal transformations of the effective  metric  
$\tilde  g_{\mu\nu}$  and  the scalar graviton $\si$ 
at an arbitrary $\ga$ as follows:
\bea\label{ga1}
{ D} \si &=& { D} \sqrt{-g}/\sqrt{-g} -  { D} \sqrt{-{\cal Q}}/\sqrt{-{\cal Q}}
=\xi^\la\partial_\la\si +2\zeta, \nn\\
 { D} \tilde  g_{\mu\nu} &=&D(e^{-\ga\si/2} g_{\mu\nu}) 
 = \tilde g_{\mu\la} \partial_\nu\xi^\la  
+ \tilde g_{\nu\la} \partial_\mu\xi^\la  +\xi^\la\partial_\la \tilde g_{\mu\nu}
+(1-\ga)\zeta  \tilde  g_{\mu\nu} \nn\\
&=& \tilde g_{\mu\la}\tilde \na_\nu \xi^\la
+  \tilde g_{\nu\la}\tilde\na_\mu \xi^\la  
+(1-\ga)\zeta  \tilde  g_{\mu\nu}, 
\eea
where $\tilde\na_\la$ means a covariant derivative 
with respect to $\tilde g_{\mu\nu}$.  Eq.~(\ref{ga1})
incorporates, in particular,  the case with $\ga=0$ at 
$\tilde  g_{\mu\nu}= g_{\mu\nu}$.\footnote{In fact, 
the transformations with the infinitesimal 
$\zeta$ may be proliferated to the finite ones  
$ g_{\mu\nu}\to Z g_{\mu\nu}$,  
$\si\to\si+2\ln Z $ and $\tilde g_{\mu\nu}  \to Z^{(1-\ga)} \tilde g_{\mu\nu}$,
with $Z\equiv e^\zeta$
being  nothing but the Weyl rescaling factor.} 

It follows  hereof that the gauge invariance of 
the Lagrangian $\tilde L_g\sim R(\tilde  g_{\mu\nu}) $  at  an arbitrary $\ga$ 
requires $\zeta\equiv 0$ leaving only the conventional GDiff. At that, inside $\tilde g_{\mu\nu}$  
there still remains one ``hidden'' extra variable.
An exception corresponds to $\ga=1$. In this case, 
$R(\tilde g_{\mu\nu}) $ describes  
the pure-tensor gravity invariant explicitly under 
GDiff times the Weyl rescaling. 
Such an extended theory
may thus be called the WGDiff gravity.   
Under WGDiff, according to (\ref{calD}) we can  first use the longitudinal Diff to achieve, say, 
the canonical value ${\cal Q}=-1$ leaving still WTDiff 
and then use the latter  (similarly to  the case with a non-dynamical ${\cal Q}$) 
to eliminate the four components from $g_{\mu\nu}$,
reducing the number of the independent components in the effective metric $\tilde g_{\mu\nu}$  
precisely to six.\footnote{The elimination 
of the longitudinal gravity component at the quantum level may, conceivably, help
in improving the quantum properties of the WGDiff 
gravity similarly to the WTDiff one~\cite{Carb3,Alv4}.}
One more  marginal case corresponds to  $\ga=0$, 
with  the metric $\tilde g_{\mu\nu}\equiv  g_{\mu\nu}$
not containing an extra variable at all and henceforth not implying the Weyl rescaling
to eliminate it. In this case,  the Lagrangian $L_g\sim R(g_{\mu\nu}) $ 
is  invariant precisely under GDiff 
representing explicitly the conventional pure-tensor gravity.
At that, according to (\ref{ga1}) the inclusion of $\si$ in an extended Lagrangian  would
explicitly restrict  the gauge symmetry 
in any case -- GDiff or WGDiff -- to GDiff as a maximal gauge 
symmetry compatible with  the scalar graviton.
Thus,  we encounter  the two conceivable patterns of the  gauge symmetry 
for the consistent  scalar-tensor theory with a dynamical~${\cal Q}$:  
GDiff ($\ga=0$) for both the  pure-tensor  and scalar-tensor cases and 
WGDiff ($\ga=1$) for the pure-tensor case, with the residual  GDiff for the scalar-tensor one.
These two alternatives are treated in detail below.

\section{Extending General Relativity}

\subsection{Basic formalism}

As a  reference case 
we choose $\ga=0$, corresponding  to GR as the prototype/graft theory, 
with  $\tilde \vphi_g\equiv 1$ and   the effective metric 
$\tilde {g}_{\mu\nu} ={g}_{\mu\nu}$
coinciding with the bare one. 
With account for the basic variations
\bea\label{deQQ}
\de\sqrt{- g}/\sqrt{-g} &=&-1/2\,{g}_{\mu\nu}\de {g}^{\mu\nu}, \nn\\
\de\sqrt{- {{\cal Q}}}/ \sqrt{-{{\cal Q}}}&=& 1/2\,{{Q}}^{-1\ka\la}\de
{{{Q}}}_{\ka\la}, 
\eea
where 
\be\label{deQQ'}
\de   {{Q}}_{\ka\la}  =  \eta_{ab}({{Q}}^a_\ka \de  {{Q}}^b_\la
+{{Q}}^{a}_\la \de {{Q}}^b_\ka), 
\ee
we then get
\bea\label{desi}
\de\si&=&\de\sqrt{-g}/\sqrt{-g}- \de\sqrt{-{\cal Q}}/\sqrt{-{\cal Q}}\nn\\
&=&-1/2( {g}_{\mu\nu}\de {g}^{\mu\nu}+{{Q}}^{-1\ka\la}\de{{Q}}_{\ka\la}).
\eea
In the above,
$ {{Q}}^{-1\ka\la}={{Q}}^{-1}{}^\ka_{a}{{Q}}^{-1}{}^\la_{b} \eta^{ab} $ 
is an inverse of ${{Q}}_{\ka\la}$, with ${{Q}}^{-1}{}^\ka_{a}\equiv
\partial x^\ka/\partial {{Q}}^a $ 
being a tetrad inverse of  $ {{Q}}_\ka^{a}\equiv\partial_\ka {{Q}}^a$.
By this token, adding $\De{\cal L}_\La$ eq.~(\ref{DeLLa}) 
to ${\cal L}_{gsm}=L_{gsm}\sqrt{-g}$~eq.~(\ref{Lgsm})  and
equating to zero the coefficients 
at the variations of the total Lagrangian density ${\cal L}_{\rm tot}$ 
with respect to  the independent
variations $\de g^{\mu\nu}$, $\de Q^a$ and $\de\phi_I$
we get the system of the field equations (FEs) in a conventional notation as follows:
\bea\label{E}
R_{\mu\nu}-\frac{1}{2}  R\, g_{\mu\nu}  -\La_g g_{\mu\nu}&=&
\frac{1}{\ka_g^2}(T_{sm\mu\nu}+\De T_{sm\mu\nu}),\nn\\
\na_\ka\Big((\ka_g^2 \La_{Q} e^{-\si}  +\frac{\de L_{sm}}{\de\si}){{Q}}^{-1}{}^\ka_a \Big)
&=&0,\nn\\
\frac{\de  L_{sm}}{\de\phi_I}  &=&0,
\eea
where $\de/\de$ means  a total  variational derivative including a derivative 
with respect to the derivative of the fields.
In the above, one conventionally has  
\be\label{ExGR'}
T_{sm\mu\nu}=\frac{2}{\sqrt{-g}}\frac{\partial (\sqrt{-g}L_{sm})}{\partial  g^{\mu\nu}}
= 2\frac{\partial L_{sm}}{\partial g^{\mu\nu}}-   L_{sm}g_{\mu\nu}
\ee
as  the canonical energy-momentum tensor of the scalar graviton and matter 
to be supplemented~by 
\be\label{DeT}
\De T_{sm\mu\nu}=- \frac{\de L_{sm}}{\de \si} g_{\mu\nu},
\ee
with $\partial/\partial$ meaning a partial variational derivative.
The second FE of (\ref{E})  restricts ultimately ${\cal Q}$ and $\si$, while
the third FE clearly accounts for matter.
Applying to the first FE of (\ref{E}) the covariant derivative 
and using the truncated Bianchi identity, 
$\na_\mu (R^{\mu\nu}-1/2\, R g^{\mu\nu})=0$, we  get 
the modified covariant conservation/continuity condition
\be\label{conserv} 
 \na_\nu (T_{sm}^{\mu\nu}+\De  T_{sm}^{\mu\nu}) =0
\ee
for the total energy-momentum tensor (but, generally,  not for $T_{sm\mu\nu}$ alone).
Stress that  $\La_Q$ does not directly enter the tensor-gravity FE (\ref{E})
(henceforth its name non-gravitating) in distinction from 
the gravitating $\La_g$ which does enter such a  FE.
Separating  
the tensor-gravity FE onto the transversal/traceless and longitudinal/trace parts 
we may present  this FE equivalently as
\bea\label{E'}
R_{\mu\nu}-\frac{1}{4}  R\, g_{\mu\nu}
- \frac{1}{\ka_g^2}(T_{sm\mu\nu}-\frac{1}{4} T_{sm}\, g_{\mu\nu})&=& 0,\nn\\
R+4\La_g+\frac{1}{\ka_g^2}(T_{sm}-4 \frac{\de L_{sm}}{\de \si}) &=& 0,
\eea
where 
\be\label{ExGR''}
T_{sm}\equiv T_{sm\mu\nu}  g^{\mu\nu}= 
2\frac{\partial L_{sm}}{\partial g^{\mu\nu}} g^{\mu\nu}-4  L_{sm}. 
\ee
The longitudinal part of (\ref{E'}) serves to restrict   
$\si$ only  modulo a scalar density~${\cal Q}$.
At that, in the neglect by the second  FE of (\ref{E}), 
signifying a non-dynamical/frozen  ${\cal Q}$, 
the latter  remains undetermined.
Under  the dynamical ${\cal Q}$,  
in  the formal limit $\La_{Q} \to \infty$  
this FE due to $e^{-\si}=\sqrt{-{\cal Q}}/\sqrt{-g}$
factorizes as $\partial_\ka( \sqrt{-{\cal Q}}{{Q}}^{-1}{}^\ka_a)=0$, 
with the (conceivably, large) CC $\La_{Q}$ getting decoupled from the classical FEs. 
Moreover, such a decoupling  takes place exactly in a specific  case of $\si$ satisfying 
the solution $\de L_{sm}/\de\si= C e^{-\si}$, with a constant $C$ (see, later). 
But generally, with the dynamical $Q^a$  making the system of   FEs dynamically closed, 
the non-gravitating CC $\La_Q$ 
survives and its manifestations  may be definite.

\subsection{Lagrange multiplier formalism}

\subsubsection{Generic case}

Basically, the independent field variables of gravity are assumed to be   
the bare metric $g_{\mu\nu}$ 
and the distinct dynamical coordinates  given by  the scalar quartet ${{Q}}^a$, 
with the (composite) $\si$ defined by~(\ref{si}).
Equivalently  (at least, on the classical level), we can use an alternative formalism 
with an  indefinite Lagrange multiplier by adding a  
constraint Lagrangian density $\De {\cal L}_\la$. 
With the latter adequately  chosen, 
such a formalism allows to make the technical procedure simpler and the physics 
interpretation  of the (formalism independent) content  more transparent.
In the case at hand,  choose the proper Lagrangian density as
\be\label{Lamult}
\De {\cal L}_\la =\la(  \sqrt{-{{{\cal Q}}}}- e^{-\si}\sqrt{-g}) 
\ee
and treat   the Lagrange multiplier 
$\la$ in the line with  the scalar-graviton field $\si$ 
as the two additional independent field variables.
Varying now the total action independently with respect 
to  $\la$, $g^{\mu\nu}$, $\si$ and ${{Q}}^a$
we   first get the relation $\sqrt{-{{\cal Q}}} =e^{-\si}\sqrt{-g}$, 
to be understood where necessary,  
followed by   FEs for the tensor and scalar gravity, 
as well as for the quartet, respectively,  as follows:
\bea\label{FEs'}
R_{\mu\nu} 
-\frac{1}{2}R g_{\mu\nu}   -\La_g g_{\mu\nu}   &=& \frac{1}{\ka_g^2}
({T}_{sm\mu\nu} +\la e^{-\si} g_{\mu\nu})  ,\nn\\
\frac{\de L_{sm}}{\de \si}+\la e^{-\si}&=&0,\nn\\
\na_\ka \Big((\la-\ka_g^2 \La_{Q}) e^{-\si}{{Q}}^{-1}{}^\ka_{a}\Big)&=&0,
\eea
with FE for matter remaining as before.
Excluding $\la$ from FEs (\ref{FEs'})  we uniquely recover  the original ($\la$-independent) 
form of FEs~(\ref{E}) 
(but  not uniquely v.v.). Such a form of the derived FEs (\ref{FEs'}),
being more transparent and  suitable  for the practical purposes, 
is  clearly due to  the appropriate  choice~(\ref{Lamult}) of~$\De {\cal L}_\la$.\footnote{We 
could   further  simplify (\ref{FEs'}) through  substitution $\la\to e^\si \la$, but this proves 
to complicate $\la$ itself (see, below).}

\subsubsection{Pure-gravity case}

Let us  now apply the preceding results to
the matterless case, $L_{sm}=L_s$, 
where the whole consideration may be executed explicitly up to the end.
Using $L_s$  from  (\ref{L_s})  at $\tilde\varphi_s=1$ 
supplemented by a scalar-graviton potential $V_s(\si)$ and accounting for  (\ref{ExGR'}) 
we now have in   (\ref {FEs'}) 
\be\label{T_s}
{T}_{s\mu\nu}=\ka_s^2\na_\mu\si \na_\nu\si -
( \frac{1}{2}\ka_s^2 g^{\ka\la}\na_\ka \si \na_\la\si  -V_s)g_{\mu\nu}
 \ee
and
\be
-\frac{\de L_{s}}{\de \si}\equiv W_s^V=
\ka_s^2\na^\la \na_\la \si +\partial V_s/\partial \si,
\ee
so that
\be
\na^\nu{T}_{s\mu\nu}=W_s^V \partial_\mu\si .
\ee
Applying  the covariant derivative to the first FE of (\ref{FEs'}) and using
the truncated Bianchi identity in combination with
the second FE of (\ref{FEs'}) 
we get that $\partial_\mu\la=0$, so that  
\be 
\la=\ka_g^2\La_0,
\ee
with $\La_0$ being an arbitrary integration constant of the dimension  mass squared. 
The scalar-graviton FE in (\ref {FEs'})  now looks like
\be\label{Ws'}
W_s^U\equiv \ka_s^2\na^\la \na_\la \si +\partial U_s/\partial \si=0,
\ee
where $U_s$ is the effective scalar-graviton potential
\be\label{Ws''}
U_s= V_s+\ka_g^2 \La_0 e^{-\si}.
\ee
Finally, the tensor-gravity FE reads  as before 
\be\label{GR'}
R_{\mu\nu} 
-\frac{1}{2}R g_{\mu\nu}   -\La_g g_{\mu\nu}  = \frac{1}{\ka_g^2}
({T}_{s\mu\nu} +\De T_{s\mu\nu}),
\ee
with the conventional energy-momentum tensor $T_{s\mu\nu}$ eq.~(\ref{T_s})
acquiring  (in compliance with (\ref{DeT})) 
the peculiar  admixture
\be\label{DeT'}
\De T_{s\mu\nu}=-\de L_s/\de \si \,g_{\mu\nu}
= \ka_g^2 \La_0 e^{-\si} g_{\mu\nu}.
\ee
At that, due to  the scalar-graviton FE (\ref{Ws'}), the total  
energy-momentum tensor of the scalar graviton (in compliance with the truncated Bianchi identity)
is  bound  to be conserved:
\be
\na^\nu(T_{s\mu\nu}+\De T_{s\mu\nu})= W_s^U\partial_\mu\si=0. 
\ee
At $\La_0=0$, we clearly recover GR extended by the CC $\La_g$ 
and a conventional scalar field~$\si$.
At  $\La_0>0$
or $\La_0<0$ we encounter two generic cases for the scalar graviton 
as a dark substance,  respectively, DE~\cite{Pir4} or DM~\cite{Pir5, Pir6}.
For completeness, 
the last FE of (\ref{FEs'}) at $\La_0-\La_{Q}\neq 0$ 
with account for $e^{-\si}=\sqrt{-{{\cal Q}}}/\sqrt{-g}$ 
factorizes~as\footnote{In the limiting case $\La_0= \La_{Q}$ 
this relation may be adopted by default.} 
\be\label{thirdeq}
\na_\ka(e^{-\si} {{Q}}^{-1}{}^\ka_{a})=
\frac{1}{\sqrt{-g}}  \partial_\ka (\sqrt{-{{\cal Q}}} {{Q}}^{-1}{}^\ka_{a})=0,
\ee
where ${\cal Q}$ is to be extracted from  $g$ and $\si$ 
due to the  the tensor-gravity and scalar-graviton FEs.\footnote{Note 
that the  previous results due to  the author, in particular,~\cite{Pir5, Pir6} 
concerning the matterless GR extension through 
the scalar-graviton  with a non-dynamical scalar density
remain, basically,  unchanged being supplemented by the quartet FE~(\ref{thirdeq}) 
to, conceivably,  restrict  the (otherwise arbitrary)  scalar density  $\cal Q$.} 
In a more general case, the solution for the Lagrange multiplier $\la$ 
may be more complicated than just a constant  (in particular, explicitly dependent on matter), 
what could make  the nature and manifestations of the emergent 
scalar-graviton dark substance more contrived.

\section{Extending Weyl transverse gravity}

\subsection{Basic formalism}

\subsubsection{Generic case}

Let now $\ga=1$ resulting in  the effective metric 
$ \tilde  g_{\mu\nu}= \hat g_{\mu\nu}$ as follows  
\be
\hat g_{\mu\nu} \equiv e^{-\si/2} g_{\mu\nu}= ({{\cal Q}}/g)^{1/4} g_{\mu\nu},
\ee
with an inverse 
$\hat g^{-1\mu\nu} =e^{\si/2} g^{\mu\nu}=(g/{{\cal Q}})^{1/4} g^{\mu\nu} $.
Such a metric is peculiar by the fact that  $\hat g =  {{\cal Q}}$
independently of the bare metric $g_{\mu\nu}$.
Now  we have for variations
\bea
\de \hat g^{-1\ka\la}
 &=&   e^{\si/2}(\de^\ka_\mu\de^\la_\nu-\frac{1}{4}  
\hat g^{-1\ka\la} \hat g_{\mu\nu})\de g^{\mu\nu}
-\frac{1}{4} \hat g^{-1\ka\la}  {{Q}}^{-1\mu\nu}\de{{{{Q}}}}_{\mu\nu},\nn\\
\de\sqrt{- \hat g}/\sqrt{\hat g} = -1/2\,{\hat g}_{\ka\la}\de {\hat g}^{-1\ka\la}&=&
1/2\,{{{Q}}}^{-1\mu\nu}\de {{{Q}}}_{\mu\nu}  
= \de\sqrt{-{{\cal Q}}}/\sqrt{-{{\cal Q}}}
\eea
as well as 
\bea\label{deg}
\de\sqrt{- g}/\sqrt{-g}&=&-1/2\, e^{\si/2}{\hat g}_{\mu\nu}\de {g}^{\mu\nu},\nn\\
\de\si&=&-1/2( e^{\si/2}\hat{g}_{\mu\nu}\de {g}^{\mu\nu}
+{{Q}}^{-1\mu\nu}\de{{{{Q}}}}_{\mu\nu}).
\eea
Adding to  $\hat{\cal L}_{gsm}=\hat {L}_{gsm}\sqrt{-\hat g}$  the CC contribution 
$\De {\cal L}_\La$ from eq.~(\ref{DeLLa}), 
equating to zero the coefficients 
at the variations of the total Lagrangian density $\hat {\cal L}_{\rm tot}$ 
with respect to  the independent
variations $\de g^{\mu\nu}$, $\de {Q}^a$ and $\de\phi_I$, 
and separating the tensor-gravity FE onto the traceless/transversal and trace/longitudinal 
respective to~$\hat g_{\mu\nu}$ parts
we get similarly to the GR case 
the system of FEs  as follows:
\bea\label{qWTDiff}
\hat R_{\mu\nu}-\frac{1}{4} \hat R\, \hat g_{\mu\nu}
- \frac{1}{\ka_g^2}(\hat T_{sm\mu\nu}
-\frac{1}{4} \hat T_{sm}\,\hat g_{\mu\nu})&=& 0,\nn\\
\frac{\de\hat L_{sm}}{\de\si}-\ka_g^2\La_g e^\si&=&0,\nn\\
\hat\na_\ka
\Big((\hat R +4\La_{Q} + \frac{1}{\ka_g^{2}}(\hat T_{sm} 
+ 4\frac{\de\hat L_{sm}}{\de\si} ) ){{Q}}^{-1}{}^\ka_a\Big)&=&0,\nn\\
\frac{\de  \hat L_{sm}}{\de\phi_I}  &=&0.
\eea
In the above, we put canonically
\be\label{hatT}
\hat T_{sm\mu\nu}\equiv \frac{2}{\sqrt{-\hat g}}
\frac{\partial (\sqrt{-\hat g}\hat L_{sm})}{\partial \hat g^{-1\mu\nu}} =
2\frac{\partial \hat L_{sm}}{\partial \hat g^{-1\mu\nu}}-
\hat L_{sm}\hat g_{\mu\nu}, 
\ee
with
\be
 \hat   T_{sm }\equiv \hat T_{sm}{}_\la^\la  
 =2\frac{\partial \hat L_{sm}}{\partial \hat g^{-1\mu\nu}} \hat g^{-1\mu\nu}
  -4\hat L_{sm}.
\ee
In  neglect (under a non-dynamical ${\cal Q}$) by the third FE of (\ref{qWTDiff}), 
the derived FEs coincide with those for the WTDiff gravity 
extended by  a specific  scalar field $\si$ (sourced by the ``gravitating'' CC $\La_g\neq 0$).
In such a case, the non-gravitating  CC $\La_Q$ 
(in the line with the scalar curvature $\hat R$) drops out from FEs  being thus  irrelevant. 
If moreover $\La_g=0$, there are  no manifestations of the vacuum energy  in FEs at all. 
Under the dynamical ${\cal Q}$,   $\La_{Q}$  
drops out only  in the formal limit $\La_{Q}\to \infty$, 
with the third FE of (\ref{qWTDiff}) factorizing in such a limit as 
$\partial_\ka(\sqrt{-{{\cal Q}}} {{Q}}^{-1}{}^\ka_a)=0$   similarly to the GR  case. 
At the large but finite $\La_Q$, the influence of the vacuum energy in the 
strong fields still survives, in contrast to the WTDiff gravity.

\subsubsection{Energy-momentum constraint}

Applying the covariant derivative to the first FE of (\ref {qWTDiff}) 
and accounting for the truncated Bianchi identity we get  
for the extended WTDiff gravity the constraint as follows:
\be\label{quasi-cont}
\frac{1}{\ka_g^2 } \hat \na^\nu   \hat T_{sm\mu\nu} =
  \frac{1}{4}\partial_\mu(\hat R  +\frac{1}{\ka_g^2}  \hat T_{sm}), 
\ee
with the  explicit dependence on the curvature $\hat R$   in distinction with the GR case.  
Due to the different behavior of the covariant conservation/continuity condition   
the two extended theories based on  GR and the WTDiff gravity are, generally,  non-equivalent. 
Still, for a particular choice of the Lagrangian $\hat L_{sm}$ (or a  specific solution  to FEs) 
which results in the fulfillment of  the condition
\be
\hat \na^\nu\hat T_{sm\mu\nu}=0
\ee
there follows from (\ref{quasi-cont}) the constraint  
\be
\hat R +\frac{1}{\ka_g^2} \hat T_{sm} =-4\La_0,
\ee
with $\La_0$ an arbitrary integration constant.
Combining such a  restriction with the first FE of  (\ref {qWTDiff}) 
we get the conventional tensor-gravity FE as follows:
\be\label{Ein'}
\hat R_{\mu\nu} -\frac{1}{2} \hat R \hat g_{\mu\nu} 
-  \La_0 \hat g_{\mu\nu}
=\frac{1}{\ka_g^2} \hat T_{sm\mu\nu},
\ee
with $\La_g$ entering only  through the second FE of (\ref{qWTDiff}) for  $\si$. 
At last, the third FE of~(\ref{qWTDiff})  now reads
\be\label{quartetFE}
\hat \na_\mu\Big((\La_{Q} -\La_0 +\La_g e^\si){{Q}}^{-1}{}^\mu_a\Big)=0
\ee
factorizing in the limit  $\La_Q\to\infty$ as before:
$\hat\na_\ka {{Q}}^{-1}{}^\ka_a
=\partial_\ka (\sqrt{-{{\cal Q}}} {{Q}}^{-1}{}^\ka_a)/ \sqrt{-{{\cal Q}}}= 0$.
Clearly, even under the assumed 
covariant conservation/continuity  
of the energy-momentum tensor of matter 
the scalar-graviton extension  (\ref{Ein'})  of WTDiff
is not completely equivalent to the similar extension~(\ref{E}) of  GR. 
Generally though,  such a conservation 
in the WTDiff case  is an additional requirement, not bound to be satisfied,
what makes the non-equivalence of the extensions to  the WTDiff 
gravity and GR even  more pronounced.\footnote{Note  nevertheless, 
that  under the covariant matter conservation  at the non-dynamical ${\cal Q}$
and  in the neglect by  the scalar graviton $\si$
the WTDiff gravity does get classically equivalent 
to GR (modulo $\La_0\leftrightarrow \La_g$).}

\subsection{Lagrange multiplier formalism}

\subsubsection{Generic case}

Let us now include  the Lagrangian density $\De\hat{\cal L}_\la$ (\ref{Lamult}) 
with a Lagrange multiplier $\hat \la$ and 
treat $\hat\la$ and $\si$  as the independent
field variables in addition to $g_{\mu\nu}$ and  ${{Q}}^a$. 
In these terms, the variations of   $\hat g^{-1\mu\nu}=e^{\si/2} g^{\mu\nu}$ 
and $\sqrt{-\hat g}=e^{-\si}\sqrt{-g}$  look like
\bea
\de\hat g^{-1\mu\nu}&=& e^{\si/2}\de g^{\mu\nu}
+\frac{1}{2}  \hat g^{-1\mu\nu}\de \si,\nn\\
\de\sqrt{- \hat g}/\sqrt{-\hat g}=-1/2\,{\hat g}_{\mu\nu}\de {\hat g}^{-1\mu\nu} &=&  
-1/2\,e^{\si/2}{\hat g}_{\mu\nu}\de g^{\mu\nu}- \de \si ,
\eea
with    $\de\sqrt{- g}/\sqrt{-g}$   given by  (\ref{deg}) 
and  $\de\sqrt{-{{\cal Q}}}/\sqrt{-{{\cal Q}}}$  
by (\ref{deQQ}) and  (\ref{deQQ'}) as before.
Extremizing the total action with respect to the independent variations $\de \hat \la$, 
$\de g^{\mu\nu}$, $\de\si$ and $\de {Q}^a$, 
we get first the constraint $e^\si=\sqrt{-g}/\sqrt{-{{\cal Q}}}$ 
and then  FEs for $\hat g_{\mu\nu}$, $\si$ and ${{Q}}^a$, respectively, as follows
\bea\label{FEs''}
\hat R_{\mu\nu} 
-\frac{1}{2}\hat R \hat g_{\mu\nu}  -\La_g e^\si    \hat g_{\mu\nu} 
 &=&  \frac{1}{\ka_g^2}(\hat {T}_{sm\mu\nu} +  \hat\la  \hat g_{\mu\nu} ),\nn\\
\hat R +\frac{1}{\ka_g^2} (\hat T_{sm} +4\frac{\de \hat L_{sm}}{\de \si}
+4\hat\la) &=&0,\nn\\
\hat \na_\ka \Big((\hat\la -\ka_g^2\La_{{Q}})
{{Q}}^{-1}{}^\ka_{a}\Big)&=&0,
\eea
with FE for matter remaining the same.
Taking trace of the first FE above and  adding up this with the second FE
we get first, in compliance  with the second FE of (\ref{qWTDiff}),   that
$\de \hat L_{sm}/\de\si = \ka_g^2\La_g e^\si$, 
and then excluding $\hat\la$ from the first two FEs of (\ref{FEs''}) we
get the tensor-gravity FE in the transversal form~(\ref{qWTDiff}).
Finally, combining the second and the third FEs above we completely recover 
the basic independent of $\hat\la$  FEs~(\ref{qWTDiff}) .

\subsubsection{Energy-momentum constraint}

Applying the covariant derivative to the tensor-gravity FE above
and accounting for the truncated Bianchi identity we get  
for the extended WTDiff gravity the constraint as follows:
\be 
\hat\na^\nu \hat T_{sm\mu\nu}=
-\partial_\mu(\hat \la+ \ka_g^2\La_g e^\si).
\ee
In the case if  $\hat T_{sm\mu\nu}$ satisfies on its own
the covariant conservation/continuity condition  
\be\label{covcons}
\hat\na^\nu \hat T_{sm\mu\nu}=0,
\ee
we get
\be
\hat \la=\ka_g^2 (\La_0 - \La_g e^\si ),
\ee
with $\La_0$ an arbitrary integration constant of the dimension mass squared. 
With the first FE of~(\ref{FEs''})  becoming now
\be\label{trace}
\hat R_{\mu\nu} 
-\frac{1}{2}\hat R \hat g_{\mu\nu}  -\La_0  \hat g_{\mu\nu}= \frac{1}{\ka_g^2}
\hat {T}_{sm\mu\nu},
\ee
the tensor gravity in this case reduces to  the (conventional) GR with a CC $\La_0$. 
Combining, in turn, the trace of (\ref{trace})  with the second FE of  (\ref{FEs''}) we  get 
the (unconventional) scalar-graviton~FE:
\be
\de \hat L_{sm}/\de\si-\ka_g^2 \La_g e^\si=0.
\ee
The remaining quartet FE clearly looks like its counterpart  (\ref{quartetFE}).
Stress ones again, that 
the covariant conservation (\ref{covcons}) 
in  the scalar-graviton WTDiff extension is an additional assumption not bound 
to be fulfilled.\footnote{Remind that a similar phenomenon, generally, 
takes place for the  scalar-graviton GR  extension where the covariant energy-momentum  
conservation is restored just 
through adding to $T_{sm\mu\nu}$ 
the term $\De T_{sm\mu\nu}$ given by~(\ref{DeT}). 
The non-conservation of the conventional energy-momentum tensor may  more generally be  
a herald of an admixture of a non-conventional  
dark substance presented  by  the scalar graviton.}

\subsubsection{Pure-gravity case}

Let us again apply the Lagrange multiplier formalism 
to the matterless  case, where the consideration may be 
proceeded up to the end, with $\hat L_{sm}=\hat L_{s}$  
given by (\ref{L_s}) at $\tilde  \vphi_s=\hat  \vphi_s=1$ 
supplemented by a scalar-graviton potential $\hat V_s(\si)$. 
First, it follows from (\ref{FEs''}) 
that the wave operator
\be\label{WLag'}
\hat  W_s^{V}\equiv - \de \hat L_s/\de\si=
\ka_s^2 \hat\na^\la\hat\na_\la \si +\partial \hat V_s/\partial\si,
 \ee
 in compliance with the second FE of (\ref{qWTDiff}),  satisfies the relation
\be\label{WLag}
\hat W_s^{V}= -\ka_g^2\La_g e^\si.
\ee
By this token, the scalar-graviton FE otherwise reads
\be\label{sgFE}
\hat W_s^U\equiv
\ka_s^2 \hat\na^\la\hat\na_\la \si +\partial \hat U_s/\partial\si   =0,
\ee
with the effective scalar-graviton potential
\be\label{sgP}
\hat U_s\equiv\hat V_s+\ka_g^2\La_g e^\si.
\ee
Likewise, we get that the energy-momentum tensor
\be
\hat {T}_{s\mu\nu}=\ka_s^2\hat \na_\mu\si \hat \na_\nu\si -
( \frac{1}{2}\ka_s^2 \hat g^{-1\ka\la}\hat\na_\ka \si \hat \na_\la\si  
-\hat V_s)\hat g_{\mu\nu}
\ee
satisfies the relation 
\be\label{W_s0}
 \hat\na^\nu \hat T_{s\mu\nu} =\hat W_s^{V}\partial_\mu\si,
\ee
implying the modified covariant conservation condition 
\be
  \hat\na^\nu (\hat T_{s\mu\nu}+\De \hat T_{s\mu\nu})=0,
\ee
with
\be\label{DeTs}
\De\hat T_{s\mu\nu}=\ka_g^2 \La_g e^\si \hat g_{\mu\nu}.
\ee
By this token, applying the truncated Bianchi identity to the first FE of (\ref{FEs''})
and integrating the result we get 
\be
\hat \la=\ka_g^2\La_0,
\ee
with $\La_0$ an arbitrary integration constant.
Henceforth, in addition to the scalar-graviton FE (\ref{sgFE}) 
we get  the tensor-gravity one as follows
\be\label{TGFE}
\hat R_{\mu\nu} 
-\frac{1}{2}\hat R \hat g_{\mu\nu}   
-\La_0  \hat g_{\mu\nu} 
=\frac{1}{\ka_g^2}
(\hat {T}_{s\mu\nu} + \De \hat {T}_{s\mu\nu}). 
\ee
For completeness, the quartet  FE of (\ref{FEs''}) at $\La_0-\La_{Q}\neq 0$ 
looks now like\footnote{At  $\La_0=\La_{Q}$ this result may be adopted by continuity.}
\be \label{quartetFE'}
\hat \na_\ka {{Q}}^{-1}{}^\ka_{a}=
  \frac{1}{\sqrt{-{\cal Q}}} \partial_\ka(\sqrt{-{{\cal Q}}}{{Q}}^{-1}{}^\ka_{a})= 0, 
\ee
similarly to (\ref{thirdeq}) for the pure-gravity extended 
GR.
At $\La_g=0$, in compliance with the ensuing covariant 
conservation of $\hat T_{s\mu\nu}$,  we recover  GR  
extended through  a conventional scalar field $\si$ 
and an emergent CC $\La_0$.

\section{Summary:  extending WTDiff  vs.\  GR   and beyond}

The quartet-metric gravity may be distinguished  by the two 
generic traits:  first, the emergence of a 
gravitational (particularly, the scalar-graviton) dark substance (such as DM, DE, etc) and, second, 
the conceivable splitting of the vacuum energy and the respective CC  onto the two parts
of the different nature --  
the gravitating $\La_g$ and the non-gravitating  $\La_{Q}$ -- 
followed by a (partial) screening of them in the dark substance environment. 
What is most  crucial, is that the  non-gravitating CC $\La_{Q}$ 
(supposed to give the dominant part of the vacuum energy)
drops out FEs in the limit  $\La_{Q}\to\infty$, getting thus suppressed 
at the large but finite values.
The latter  phenomenon seems to be typical 
within  the quartet-metric frameworks irrespective of the particular realization mode. 
On the other hand, the behavior of a remaining gravitating part of the vacuum energy 
may depend significantly on a realization mode.  
The quartet-metric paradigm being extremely rich in its 
prospects for going beyond GR, possesses, by the same  token, by many ambiguities.
One is in choosing between the  alternative prototype/graft theories -- GR or  WTDiff --
on which the extension should be built, followed by the evident ambiguity in constructing 
a particular effective Lagrangian.
This is due to the fact that  though the given  prototype theories are (under the proper assumptions) 
classically equivalent their scalar-graviton extensions ceases, generally,  to be such. 

This non-equivalence  was explicitly demonstrated in the 
pure-gravity case 
under the simplest choice of the scalar-graviton Lagrangian,
what allows  the consideration  to be executed up to the~end. 
More particularly, in the GR extension 
the   Lagrangian gravitating CC $\La_g$, supposed to be subdominant,  
influences the tensor-gravity FE  (\ref{GR'})  directly,
signifying an ``inborn'' CC problem as in GR itself. 
At that, the  scalar-graviton enters this FE through the 
admixture (\ref{DeT'})
to the canonical scalar-graviton energy-momentum tensor, 
with such an admixture being proportional to an arbitrary integration constant $\La_0$ 
(defining ultimately the kind of the emergent dark substance: DM, DE, etc).
On the contrary, in  the WTDiff extension  the tensor-gravity FE~(\ref{TGFE}) 
contains a similar integration constant $\La_0$ as an emergent  CC, 
what may smother the quantum behavior of the latter  compared to the Lagrangian CC $\La_g$ 
in the GR extension. At the same time, 
$\La_g$ enters the tensor-gravity FE in the WTDiff extension 
only  through the admixture (\ref{DeTs}) to 
the canonical energy-momentum tensor of the scalar graviton, 
with the behavior of such an admixture  dependent on the solution 
to FEs.\footnote{It may be noted that under  minimally extending  GR and WTDiff 
the Lagrangian and emergent CCs, respectively,  $\La_g$ and $\La_0$  
in the tensor-gravity FE interchange each other 
(supplemented by $\si\leftrightarrow  -\si$). 
This partially explains the difference between the extensions.}  
This may result  in the WTDiff extension in an additional (besides that for $\La_Q$) 
screening of the  vacuum energy paving,  conceivably,  
the way to consistently solving here the CC problem.
Clearly, this requires further investigation.

To conclude,  following comparatively both these routes 
may, hopefully, shed more light on the scalar graviton as a kind of  an astroparticle,
as well as, more generally, on the quartet-metric paradigm as a whole.

\paragraph{Acknowledgment}
The author is grateful to S.S.\ Gershtein for  interest to the work and  encouragement.

\end{document}